\documentclass{ieeeaccess}
\usepackage{cite}
\usepackage{amsmath,amssymb,amsfonts}
\usepackage{algorithm}
\usepackage{algpseudocode}
\usepackage{graphicx}
\usepackage{textcomp}
\usepackage{xcolor, soul}
\sethlcolor{yellow}

\def\BibTeX{{\rm B\kern-.05em{\sc i\kern-.025em b}\kern-.08em
    T\kern-.1667em\lower.7ex\hbox{E}\kern-.125emX}}
\begin{document}
\history{Date of publication xxxx 00, 0000, date of current version xxxx 00, 0000.}
\doi{10.1109/ACCESS.2023.0322000}

\title{Efficient UAVs Deployment and Resource Allocation in UAV-Relay Assisted Public Safety Networks for Video Transmission}
\author{\uppercase{Naveed Khan}\authorrefmark{1},
\uppercase{Ayaz Ahmad}\authorrefmark{1}, \uppercase{Abdul Wakeel}\authorrefmark{2}, \uppercase{Zeeshan Kaleem}\authorrefmark{1}, \uppercase{Bushra Rashid}\authorrefmark{3}, and \uppercase{Waqas Khalid}\authorrefmark{4}}

\address[1]{Department of Electrical and Computer Engineering, COMSATS University, Islamabad, Wah Campus, Pakistan}
\address[2]{Electrical Engineering Department, Military College of Signals (MCS), National University of Sciences and Technology (NUST), Islamabad, Pakistan}
\address[3]{Department of Biomedical Engineering, College of Engineering, King Faisal University, Alahsa, Kingdom of Saudi Arabia}
\address[4]{Institute of Industrial Technology, Korea University, Sejong 30019, South Korea}

\corresp{Corresponding authors: Waqas Khalid (waqas283@\{korea.ac.kr, gmail.com\}) and Ayaz Ahmad (ayaz.uet@gmail.com)}

\tfootnote{This research was supported by Basic Science Research Program through the National Research Foundation of Korea (NRF) funded by the Ministry of Education(MOE) (NRF-2022R1I1A1A01071807).}

\begin{abstract}
Wireless communication highly depends on the cellular ground base station (GBS). A failure of the cellular GBS, fully or partially, during natural or man-made disasters creates a communication gap in the disaster-affected areas. In such situations, public safety communication (PSC) can significantly save the national infrastructure, property, and lives. Throughout emergencies, the PSC can provide mission-critical communication and video transmission services in the affected area. Unmanned aerial vehicles (UAVs) as flying base stations (UAV-BSs) are particularly suitable for PSC services as they are flexible, mobile, and easily deployable. This manuscript considers a multi-UAV-assisted PSC network with an observational UAV receiving videos from the affected area's ground users (AGUs) and transmitting them to the nearby GBS via a relay UAV. The objective of the proposed study is to maximize the average utility of the video streams generated by the AGUs upon reaching the GBS. This is achieved by optimizing the positions of the observational and relay UAVs, as well as the distribution of communication resources, such as bandwidth, and transmit power, while satisfying the system-designed constraints, such as transmission rate, rate outage probability, transmit power budget, and available bandwidth. To this end, a joint UAVs placement and resource allocation problem is mathematically formulated.  The proposed problem poses a significant challenge for a solution. Considering the block coordinate descent and successive convex approximation techniques, an efficient iterative algorithm is proposed. Finally, simulation results are provided which show that our proposed approach outperforms the existing methods.

\end{abstract}
\titlepgskip=-21pt
\maketitle
\section{Introduction}
\label{sec:introduction}
\PARstart{I}{n} mobile communications, cellular base stations may fail due to natural or man-made disasters. Additionally, deploying advanced networks in some areas can be impractical or too risky for first responders. For example, the Sichuan (China) earthquake, in May 2008, caused significant damage to telecommunication infrastructure \cite{1}. Similarly, the World Trade Center attack on 9/11 led to nearly 3 million users losing cell phone services \cite{2}. In such situations, individuals require guidance, counsel, and real-time information about ongoing situations. Often, they must send requests for help or share their locations with rescue teams. In disaster settings, where cellular networks often fail, it is particularly beneficial for users to establish direct communication with each other \cite{2a}. To address these challenges, using an unmanned aerial vehicle (UAV) as a cellular base station (BS) is a key solution that allows fast, seamless, and reliable cellular communication, crucial for public safety \cite{3}. UAVs do not need any restricted and costly calibrations, e.g., cables, and can effortlessly move and dynamically adjust their positions, to yield on-demand cellular communications to affected area ground users (AGUs) in undesirable circumstances\cite{4}.
Moreover, UAVs offer a compelling solution for enhancing the functionality of wireless networks in boasting various applications, including offloading wireless video services in regions lacking proper infrastructure or during disasters \cite{4a}.

The demand for mobile video streaming in everyday life as well as in rescue operations is on the rise, dominating global mobile data traffic. As per the Ericsson Mobility Report for 2022 \cite{4b}, video streaming forms a substantial and swiftly expanding segment of mobile data traffic. In 2022, video traffic represented 70\% of all cellular data, and it is expected to grow to 80\% in the coming years. However, this surge in video streaming popularity is causing congestion issues, particularly for users at the edge of cellular coverage, leading to a degradation in the quality of experience (QoE) for these end users. To address these challenges, recent research efforts have been focused on developing efficient solutions to meet users' QoE requirements. One conventional approach involves deploying small-cell networks comprising numerous small base stations (SBSs) \cite{4b1}. However, this strategy may prove cost-ineffective in scenarios with temporary spikes in user density or highly dynamic traffic demands. Alternatively, UAVs have emerged as a viable alternative for extending wireless network coverage and alleviating congestion issues. UAVs can offload traffic and provide video services, particularly in areas affected by disasters \cite{4b2}. Furthermore, compared to direct transmissions from terrestrial BS over long distances, the use of UAVs as relays offers several advantages, such as the potential for line-of-sight (LoS) links with adjustable altitudes, expanded coverage for ground wireless devices, and reliable uplink and downlink connections for affected ground users \cite{4f}. 

UAVs, when equipped with wireless transceivers, can transmit video footage from the affected area during disasters. Generally, the effectiveness of these wireless transceivers is contingent upon the size of the antenna they employ. Utilizing high-performance larger antennas can enable UAVs to transmit data over larger distances. However, due to the constrained payload capacities of UAVs, high-performance antennas tend to be too large and heavy for them to carry while maintaining flight capabilities. Consequently, when a single UAV is tasked with observing and transmitting real-time video data, the range of communication is restricted, thus limiting the scope of real-time observation. Nevertheless, this constraint can be addressed through the use of multiple UAVs, which can establish communication with one another via their integrated wireless transceivers. This interconnected arrangement is commonly referred to as a UAV network \cite{4g}.  \par
In this paper, we propose a public safety communication network (PSCN) consisting of observation UAV and relay UAV for uplink video streaming in a fading channel wireless environment. The network utilizes an observation UAV as an aerial base station to receive video streams from AGUs and transmit them to the nearby functional ground base station (GBS). Simultaneously, the relay UAV bridges the connection between the observation UAV and the GBS while extending the coverage of PSCN.  Designing such a UAV-based uplink video streaming system requires careful consideration of parameters, including the aerial placement of observation and relay UAVs, as well as the allocation of resources such as transmission power and bandwidth. These factors are crucial for achieving superior signal strength and ensuring equitable delivery of good quality video for all AGUs \cite{4h}.

\subsection{Related Work}
\label{literaturereview}
UAVs play a pivotal role in cellular communications, particularly in enhancing public safety operations. Despite their significant impact, UAV deployment faces key challenges, including energy limitation \cite{5}, flight time constraints, optimal trajectory \cite{7}, interference and resource management \cite{9}, and efficient placement. Addressing these issues is crucial for the efficient utilization of UAVs in public safety scenarios. Different approaches have been proposed for the optimal deployment of UAVs, focusing on minimizing UAV transmit power, expanding wireless coverage, and leveraging UAVs for relay and video streaming services. In the subsequent sections, we discuss some of the well-known approaches in this context.

\subsubsection{UAV Placement and Recourse Allocation}
\label{sub:UAVplace}
In \cite{11}, the authors have proposed a single UAV-based disaster management system for indoor employees in a multi-storeyed building to deliver wireless coverage. The authors have considered the problem of efficient UAV deployment to cover the entire building with minimum transmit power.   
The authors in \cite{14} have applied the Particle Swarm Optimization algorithm in an indoor scenario. The objective of their study is to optimize the 3-D placement of UAVs to minimize the overall transmit power. By utilizing the Particle Swarm Optimization algorithm, the authors aim to find an optimal configuration that reduces the total power consumption required for wireless communication in the indoor environment.\par
The authors in \cite{15} have proposed an optimal UAV placement system for delivering cellular services to outdoor and indoor users bypassing the load of the existing wireless network infrastructure with low transmit power. In \cite{15a}, the authors introduced a reinforcement learning approach to address the challenge of enhancing the max-min sum rate in a computationally demanding scenario by determining the optimal 3D UAV (ABS) positions and power allocation
In \cite{16}, Shukla et al. analyzed the resource distribution of multiple UAV-based systems subject to minimizing the energy consumption and operation delay taking into account the edge and cloud servers. In \cite{17}, Zhaohui et al. have minimized the sum power of the UAVs by jointly considering the beam width, height, bandwidth allocation, and position of the UAVs as optimization variables. in \cite{17a}, the authors delved into the joint optimization of resource allocation and relay selection within a decode-and-forward downlink OFDMA cooperative network, considering the scenario of outdated CSI at the base station.
The work in \cite{18} has analytically assessed the rotational angle division multiple access (RADMA) performance implemented using UAV-BS with array antennas. To demonstrate how RADMA performs better in a low-power wide-area network configuration, a numerical performance evaluation is carried out. According to the simulation results, RADMA can reduce the transmission energy needed by 20\%, transmission time by 25\%, and the packet loss ratio by 77\%.\par

Besides minimizing the transmit power, the focus is also to maximize the cellular coverage of the UAVs for serving a large number of users. The authors in \cite{19} formulated a 3-D UAV placement problem with the objective of cellular coverage maximization. The authors in \cite{20} have considered the optimal placement of UAVs using a circle-packing algorithm. They have demonstrated that the cellular coverage is maximized by fixing the UAVs 3-D location. User-centric and network-centric approaches were introduced by Kalantari et al. in \cite{21}. For each of their approaches, the authors determined 3-D coordinates for optimal deployment of the UAVs that maximize the sum rate and cellular coverage.\par 
The authors in \cite{22} introduced a strategy for positioning multiple UAV-based base stations in a manner that maximizes cellular access for users while minimizing inter-symbol interference. Two methods were proposed for the UAVs placement. In the first method, a linear algorithm was Suggested for the sequential deployment of the UAVs. In the second method, the authors used concurrent UAV placement with machine learning algorithms. The experimental results obtained show that user coverage can be enhanced if the ground users are spread unevenly. The authors in \cite{23} have maximized the coverage area of UAV-based cellular communication networks subject to UAV hovering duration and the average sum of bits transmitted to the users. In \cite{24}, the authors analyzed the potential of deploying multiple UAVs to improve coverage performance when interference is a factor. The proposed approach suggests that by adjusting the height of the UAVs in different working environments, it is possible to optimize coverage and determine the best configuration for a particular set of UAVs. This method provides valuable design guidelines for achieving better coverage and minimizing interference among UAVs. \par 
In \cite{26}, the authors have studied the placement of UAVs in a catastrophe-affected zone to handle the high influx of user traffic that typically occurs during emergencies. The authors aim to identify the optimal UAV coverage that would ensure the maximum user throughput with fair distribution across all areas of the topology. The authors in \cite{27} deployed UAVs in a square-shaped area, assuming that all links within the region have Line-of-Sight (LoS) and directional antennas are utilized to maximize coverage.

An attractive area of research is to deploy UAVs for video streaming. A flexible amount of support is provided to the existing cellular networks by deploying UAVs for multimedia services in real-time scenarios.\cite{27a} proposed a virtual reality (VR) wireless network, where the authors used UAVs to transmit the acquired video data to a ground BS. The authors in \cite{27a1} investigated the power allocation and video bitstream adaptation in the context of video streaming across multi-node wireless networks where interference varies over time.
\cite{27b} and \cite{27c} have analyzed energy efficiency in a UAV-enabled video streaming scenario by keeping the flight height of different UAVs fixed. For multi-user video transmission, the authors in \cite{27d} optimized bandwidth and transmit power distribution in a UAV relay network to maximize the long-term QoE of the users.\par
In \cite{27e}, the authors have explored UAV video streaming in sensor-augmented systems and proposed video streaming algorithms for the sensor data. In \cite{27f}, the authors proposed an optimized heterogeneous framework for a video cellular network, where the authors have jointly optimized the rate allocation, selection of video quality, and time-domain resource portioning. In \cite{27g}, a live video streaming forest fire surveillance system was proposed. The system is low-cost to implement, simple to use, and has a large coverage area. For adaptive streaming in mobile wireless networks, utility maximization was proposed in \cite{27h}. The authors have shown that maximum utility can be achieved through efficient resource allocation and dynamic rate adaptability.

UAVs typically rely on limited power sources, primarily batteries, which constrain their flight duration. For instance, the majority of commercial rotary-wing UAVs have a maximum flight time of just around 30 minutes \cite{10a}. Consequently, various strategies have been proposed to extend UAV flight durations. These solutions include battery swapping \cite{10b}, harnessing solar power for charging \cite{10c}, RF-based charging \cite{10d}, laser-based wireless power transfer (WPT) \cite{10e}, and optimizing UAV deployment \cite{10f}. \cite{10g} presented a 750-W 85-kHz band inductive rapid charging system designed for mid-sized UAVs (drones). This system was intended for opportunistic charging during continuous industrial operations. Additionally, in \cite{10h}, the authors explored the adoption of renewable energy generation and storage equipment integrated with traditional charging stations. The aim is to minimize reliance on power sourced from the distribution network. In \cite{10i}, authors examined the concept of drone-to-drone opportunity charging as a means to enhance flight duration, particularly in the context of multi-agent systems.
\subsubsection{UAV Deployment as a Relay}
\label{UAVRelay}
Recently, different authors have worked on the deployment of UAVs as relays over a region, to deliver cellular coverage to the ground users when a direct communication link is blocked either due to malfunctioning of the existing base station or long distance from the BSs. In \cite{28}, the authors have considered multiple pairs of ground user scenarios where direct communication links are not available either due to long distances or no LOS, i.e., blockages. For multiple pairs of ground users, the authors deployed a single UAV as a relay in a time-division manner, and the minimum average rate is maximized by allocating optimum time slots to all communication pairs. In \cite{29}, the authors have investigated an optimal 3-D placement of UAVs as relays when the existing base station malfunctions. Three different models were used to achieve optimal 3-D placement of the relay UAV. In \cite{30}, the authors have proposed a matching theory to analyze the advantages and difficulties of UAVs as relay models in bulky UAV communication systems. \par
The authors in \cite{31} have proposed UAV-based cellular communication systems in which a relay-based UAV is utilized for bridging a set of distant user equipment (UE) with the ground base station and vice versa. This two-way communication is possible when the UAV is relayed by orthogonal frequency (OF) bands. The UAVs transmission power and placement are jointly optimized to achieve a high total rate for both down-link and up-link communication subject to the SNR constraints and communication power constraints on the UAV control channel. In \cite{32}, the authors have studied UAV as a relay for uplink data transmission to connect as many IoT devices as possible while still meeting their heterogeneous quality of service needs. In \cite{33}, the authors worked on UAV placing problems in a relay network by integrating the local topological knowledge, where the aim was to deploy the UAV in a location with a good LOS. The authors in \cite{34} proposed dynamic positioning systems to build a communication link between the disjoint nodes by utilizing UAV as a relay. The authors suggest that their approach could find practical applications in real-life situations, such as during natural disasters like floods and earthquakes, as well as to gather data from deployed sensor nodes. In \cite{35}, the authors introduced a UAV-based emergency Wi-Fi system to assist the rescue activities by supervising the survivors at the nearby rescue camp, whereas, the authors in \cite{36} have designed UAV-based energy-efficient relay systems that jointly optimize the BS and UAV transmit power.
\subsection{Contributions}
\label{Contributions}
Different from the aforementioned research, herein, we analyze joint optimization of the observation and relay UAVs locations, transmit power, and bandwidth distribution for different numbers of AGUs over a fading channel in PSCNs. The main objective is to augment the average utility of video streams created by the AGUs when arriving at the GBS over a fading channel. The primary contributions of this manuscript can be summarised as follows:
\begin{itemize}
\item We propose a multi-UAV-based up-link video streaming system for video transmission in fading channel conditions within PSCNs. Analytical expressions for the rate outage probability between the observation UAV and AGU over the fading channel are calculated.
Furthermore, we formulate a video streaming utility function for each AGU. Subsequently, we develop an optimization problem aimed at maximizing the average streaming utility for all AGUs. The optimization problem involves a joint optimization of the observation and relay UAV's locations, transmit power, and allocation of the bandwidth. The optimization is subject to the information causality constraint of the relay UAV and GBS, as well as the outage probability constraint between the observation UAV and AGUs.
\item  Due to the complexity of the formulated problem, a direct solution is challenging. Consequently, an iterative algorithm is proposed to obtain an efficient solution. This algorithm utilizes successive convex approximation (SCA) and block coordinate descent techniques. In particular, two auxiliary problems are optimized iteratively: the allocation of bandwidth and power with fixed UAV locations, and the positioning of UAVs with fixed bandwidth and transmit power.
\item We present simulation results to demonstrate the trade-off between the data rate of the observation UAV and the relay UAV, as well as the trade-off between the relay UAV and the GBS. Additionally, we provide evidence of the effectiveness of our proposed design by evaluating the average streaming utility.
\end{itemize}
\par
The remaining manuscript is organized as follows: Section II provides the problem formulation and system model of our proposed PSCNs, whereas, the mathematical solution of the formulated problem is given in Section III. Section IV presents the system parameters, numerical findings, and simulation results. Concluding remarks and future research directions are provided in Section V.
\section{System model and problem formulation}
\label{sysmodelandps}
Herein, we present the problem statement and a UAV-based video streaming PSCNs system model. Table~\ref{table_1} summarizes the important notations used and the explanations that go with them.
\begin{table}[h!]
\caption{Parameters definition}
\label{table_1}
\begin{center}
\begin{tabular}{|l|l|}
\hline
\textbf{Notation} & \textbf{Definition}  \\
\hline
B & System bandwidth ($Hz$) \\
\hline
$H_o$ &  Observation UAV elevation ($m$)\\
\hline
$H_b$ &  Ground BS elevation ($m$) \\
\hline
$U$ & Total number of affected ground users (AGUs)\\
\hline
$P_o^{max}$ & Observation UAV maximum transmit power (Watts) \\
\hline
$P_r^{max}$	& Relay UAV maximum transmit power (Watts)\\
\hline
$P_u^{max}$	& Affected area users' maximum transmit power (Watts)\\
\hline
$n_u$ & $u$th affected ground user(AGU) \\
\hline
$w_b$ & Ground BS's two-dimensional (2D) coordinates \\
\hline
$w_u$ & 2D coordinates of $n_u$\\
\hline
$q_o$ &  2D coordinates of Observation UAV  \\
\hline
$q_r$ &	2D coordinates of Relay UAV \\
\hline
$\rho$	& Target outage probability threshold \\
\hline
$x_u$ & Part of the total bandwidth allocated to $n_u$ \\
\hline
$P_u^{out}$ & Rate outage probability of user $n_u$\\
\hline
$P_u$ & Transmit power assigned to the $n_u$ user \\
\hline
$R_u$ & Uplink transmit rate over AGU to observation \\
& UAV link for AGU $n_u$ (bps/Hz)\\
\hline
$R_r$ & Transmit rate over observation UAV \\
& to relay UAV link (bps/Hz)\\
\hline
$R_b$ & Transmission rate of relay UAV to \\
&  the ground BS (bps/Hz)\\
\hline
$D$ &  D represent the PSCNs size\\
\hline
$\overline{r}_u$ & Required video playback rate (bps) of AGU $n_u$\\
\hline
$S_u$ & AGU $n_u$ streaming utility\\
\hline
\end{tabular}
\end{center}
\end{table}
\subsection{System model}
\label{sysmodel}
\begin{figure}[ht]
\centering
\includegraphics[width=0.8\linewidth]{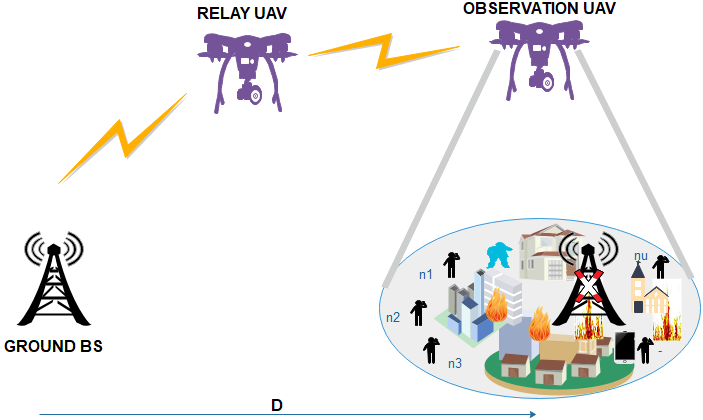}
\caption{System overview: Overall geographical scenario}
 \label{fig1}
\end{figure}

In our proposed PSCN, we consider an up-link UAV-based video streaming system where AGUs are connected to the GBS through observation and relay UAVs. These UAVs are utilized to offload the video streams from the AGUs to the GBS over a fading channel, as illustrated in Fig.~\ref{fig1}. In our proposed PSCN, we employ one observation UAV and one relay UAV both performing different tasks. The observation UAV is primarily used to receive and re-transmit the video generated by the AGUs, while the relay UAV serves as the connection between the observation UAV and the GBS for real-time video transmission over longer distances. The observation UAV is not suitable for relaying data between AGUs and the GBS due to its long distance from the GBS. Furthermore, being far away from the rely UAV, the AGUs cannot establish direct communication with it.

In Fig.~\ref{fig1}, the group of AGUs is represented by ${U} = \{n_1, n_2,\cdots, n_u\}$, where the observation UAV equipped with a wireless transceiver is used to receive the video streaming services from these AGUs. Moreover, the positions of the AGUs are assumed to be known. In a three-dimensional (3-D) Cartesian coordinate system, the coordinates of the observation UAV and the relay UAV are represented as $(a_o, c_o, H_o)$ and $(a_r, c_r, H_r)$, respectively. Moreover, the coordinate of the GBS is represented as $(a_b, c_b, H_b)$. Likewise, $(a_u, c_u, 0)$ are the coordinates of the AGU $n_u, 1 \leq u \leq U$. We further define $w_b~=~[a_b, c_b]^T$ and $w_u~=~[a_u,~c_u]^T$ to depict the aforementioned 2-D point projection on the ground plane.\par
We suppose that the observation and relay UAVs are used for the provisioning of communication services under the communication range/coverage of a single base station, therefore, UAVs mobility is not required. That is, only the optimal placement of both UAVs is considered. Both the observation and relay UAVs are always located at a set elevation above the ground, symbolized by $H_o$ and $H_r$, respectively. Furthermore, D in Fig.\ref{fig1} represents the distance from the center of the observation area to the GBS. Let the ground plane coordinates of the observation and relay UAVs be represented by $q_o = [a_o, c_o]^T \in R^{(2\times1)}$ and $q_r = [a_r,c_r]^T \in R^{(2\times1)}$, respectively. Based on these positions, we can mathematically express the distances between the AGU $n_{u}$ and the observation UAV $d_{uo}$, between the observation UAV and the relay UAV $d_{or}$, and between the relay UAV and the GBS $d_{rb}$ as follows
\begin{eqnarray}
d_{uo}&=&\sqrt{(H_o)^2~+~\parallel~q_o~-~w_u\parallel^2}\nonumber\\
d_{or}&=&\sqrt{(H_r - H_o)^2 + \parallel q_r - q_o \parallel^2}, \qquad \mbox{and}\nonumber\\
d_{rb}&=&\sqrt{(H_b - H_r)^2 + \parallel w_b - q_r\parallel^2} \nonumber \;,
\end{eqnarray}
respectively. 
\subsection{Channel model}
\label{chanelmodl} 
The communication link between AGUs and the observation UAV is dominated by large-scale and small-scale fading \cite{27b} whereas the channel between the observation and relay UAVs and the relay UAV and GBS are dominated by the free-space path loss (FSPL) model. The fading channel coefficient between the AGU $n_{u}$ and observation UAV can be modeled as $h_{uo} = \epsilon_u \sqrt{\gamma_u }$, where $\epsilon_u$ is a complex random variable that accounts for small-scale fading, whereas the large-scale channel attenuation between the observation UAV and the AGUs is represented as ${\gamma_u}$. Specifically, with $d_{uo}$ been the distance between the observation UAV and the AGU $n_{u}$,then
\begin{eqnarray}
\gamma_u & =& \frac{\alpha_0}{(d_{uo})^2} \nonumber \\
& = &\frac{\alpha_0}{\left(\sqrt{(H_o)^2 + \parallel q_o - w_u \parallel^2}\right)^2}\;, \quad \forall u
\label{eq:1}
\end{eqnarray}
where $\alpha_0$ represent the channel gain at $d_0 = 1m$. The small-scale fading among the AGUs and observation UAV follows the Rician fading model as the AGUs and observation UAV are in Line-of-Sight (LoS) \cite{46}. The Rician factor $K_C$ reflects the power ratio between the LoS signal constituent and the dispersed constituents in the Rician fading channel. Consequently, the CDF of $|\epsilon_u|^2$ can be written as \cite{27b, 46},
\begin{eqnarray}
F(z) & \cong & p(|\epsilon_u |^2 < z) \nonumber \\
& = &1 - Q_1 \left(\sqrt{2K_C}, \sqrt{2(K_C + 1) z}\right)\;,
\label{eq:2}
\end{eqnarray}
where $Q_1(\cdot)$ represent the Marcum-Q function. The Marcum-Q function with modified Bessel function $I_0$ of order 0 represented as $Q_1(a,b)$ can be written as 
\begin{equation}
\label{eq:3}
Q_1(a, b) =\int_b^{\infty}x~ e^{\frac{-x^2 + a^2}{2}}~I_0 (ax)~dx\;.
\end{equation}
Let $P_u$ be the user's transmit power and $N_0^2$ be the noise power density at the AGU, the channel capacity (b/s/Hz) between the AGUs and observation UAV can then be defined as
\begin{eqnarray}
\label{eq:4}
C_u & = \log_2\left(1 +\frac{ P_u |h_{uo}|^2}{N_o^2}\right)\nonumber \\
&= \log_2\left( 1 +\frac{P_u \gamma_u |\epsilon_u|^2}{N_o^2}\right)
\end{eqnarray}
For simplicity, in our proposed PSCNs, we allocate equal bandwidth to both the information received by the observation UAV from the AGUs and its transmission to the GBS. Furthermore, we adopt FDMA as the communication technique, distributing the total bandwidth $B$ among all the AGUs. Precisely, designating $x_u$ as the fractional bandwidth allotted to users $n_{u}, x_u \geq 0$ and $\sum_u^U x_u \le 1$. Furthermore, FDMA is chosen for interference mitigation, easy implementation, and spectrum efficiency \cite{56}. The normalized attainable rate (b/s/Hz) for the AGUs uplink communication can be calculated as

\begin{eqnarray}\label{eq:5}
R_u &=& x_u\log_2\left(1+\frac{P_u|h_{uo}|^2}{x_u B N_o}
\right) \nonumber \\
&=& x_u \log_2\left( 1 +\frac{P_u\mu_0 |\epsilon_u|^2}{x_u((H_o )^2 + \parallel q_o - w_u \parallel^2)}\right),
\end{eqnarray}
where $\mu_0 = \frac{\alpha_0}{B N_o}$. We define the rate outage probability $P_u^{out}$ as the probability that the communication between the AGU and the observation UAV fails, written as
\begin{eqnarray}
\label{eq:6}
p_u^{out} & \triangleq & P(C_u < R_u) \nonumber \\
& = & P\left(|\epsilon_u|^2 < \frac{2^{R_u / x_u -1}}{P_u \gamma_u /N_o^2} \right) \nonumber \\
& = & F\left( \frac{(2^{R_u/x_u -1})((H_o)^2 +\parallel q_o - w_u \parallel^2)}{P_u \mu_0}\right)
\end{eqnarray}
\par
Due to the CDF definition in Eq.~\ref{eq:2}, the last equality holds. Subsequently, we indicate the outage probability as the rate outage probability as expressed in Eq.~\ref{eq:6}. Furthermore, it is assumed that the channels between the observation UAV and relay UAV, as well as the relay UAV and GBS, are primarily characterized by LoS propagation. As a result, the channel gains for these two links can be effectively modeled using the FSPL equation.
\begin{equation}
\label{eq:7}
h_{or} = \frac{\alpha_0}{(d_{or})^2}\;, \quad \mbox{and} \quad h_{rb} = \frac{\alpha_0}{(d_{rb})^2}\;,
\end{equation}
Let $P_o$ denote as the transmit power of the observation UAV, then the normalized attainable rate of the relay UAV (measured in bits per second hertz, b/s/Hz) can be calculated as
\begin{eqnarray}
\label{eq:8}
R_r & = & \log_2\left( 1 + \frac{P_o ~ h_{or}}{B N_o}\right) \nonumber \\
& = &\log_2\left(1 + \frac{P_o~\mu_0}{(H_r - H_o)^2 + \parallel q_r - q_o\parallel^2}\right)
\end{eqnarray}
Similarly, let $P_r$ be the relay UAV transmit power, then the attainable rate of the ground BS (b/s/Hz) can be written as
\begin{eqnarray}
\label{eq:9}
R_b & = & \log_2\left(1 + \frac{P_r~h_{rb}}{B N_o}\right) \nonumber\\
& = & \log_2\left(1 + \frac{P_r~\mu_0}{(H_b - H_r)^2 + \parallel w_b - q_r\parallel^2}\right)
\end{eqnarray}
\subsection{Video streaming model}
\label{adaptiveVSM}
For video transmission from the AGUs, we assume the use of adaptive video streaming. To model the utility of each user, we adopt a simple HTTP video streaming utility model that solely depends on the video transmission rate. Specifically, when the GBS receives a higher video rate from an AGU, it perceives a correspondingly higher video quality for that particular AGU \cite{47}. Moreover, we define the video streaming utility paradigm based on a logarithmic relationship with the video transmission rate, i.e., $S_{u}~=~\theta \log (\beta R_u/ \overline{r}_u)$ as in \cite{48, 49}. The positive constant factors $\theta$ and $\beta$ vary for different sorts of applications. $\overline{r}_u$ represents the playback rate for AGU $n_{u}$, which is related to the media outlet's physical capabilities (e.g., users' data rate and screen size). Therefore, the AGUs need to use an appropriate playback rate that matches the video's frame rate and resolution to ensure smooth and accurate playback at the BS. If the playback rate is too high or too low, the video may appear choppy, stutter, or be out of sync with the audio. The average video streaming utility $\overline{S}$ for all the AGUs can be written as
\begin{equation}
\label{eq:10}
\overline{S} = \frac{\theta}{U} \sum_u^U \log\left(\beta \frac{R_u}{\overline{r}_u} \right)\;.
\end{equation}
It is noteworthy to mention that $\overline{S}$ is a positive function that is concave for $R_u$, continuous, and differentiable w.r.t $R_u$.
\subsection{Problem formulation}
\label{probformu}
Lets $\mathcal{X} = \{x_u, \forall u\}$, $\mathcal{P} = \{P_u, P_o, P_r, \forall u\}$, and $\mathcal{Q} = \{q_o, q_r\}$.The goal is to maximize the average video streaming utility for all the AGUs in a UAV-based PSCN by jointly optimizing the positions of the observation and relay UAVs, along with power allocation $\mathcal{P}$ and bandwidth distribution $\mathcal{X}$. This optimization is subject to the constraints of information causalities at the relay UAV and GBS, as well as the outage probability constraint for the channel between the observation UAV and AGUs, which should be kept below a specified threshold $\rho$. The problem is reformulated as 
\begin{equation}
\mathcal{P}1: \begin{array}{c}
    max  \\
     \mathcal{X}, \mathcal{P}, \mathcal{Q}
\end{array}\frac{\theta}{U} \sum_u^U\log \left(\beta \frac{R_u (1-\rho)}{\overline{r}_u} \right)
\nonumber
\end{equation}
$\qquad \qquad \qquad \qquad \qquad \mbox{\textbf{s. t.}}$
\begin{eqnarray}
\label{eq:11}
P_u^{out} & \leq & \rho \quad \forall~u, \\
\label{eq:12}
R_r & \geq & \sum_u^U~R_u \\ 
\label{eq:13}
R_b & \geq & R_r \\
\label{eq:14}
\sum_u^U~x_u & \leq & 1 \\
\label{eq:15}
0 \leq & x_u  &\leq 1, \qquad \forall u \\
\label{eq:16}
0 \leq & P_u \leq & P_u^{max}\;, \qquad \forall u\\
\label{eq:17}
0 \leq & P_o \leq & P_o^{max}\; \qquad\\
\label{eq:18}
0 \leq & P_r\leq & P_r^{max}\; \qquad 
\end{eqnarray}
In the formulation of the above problem, constraint (\ref{eq:12}) and (\ref{eq:13}) are information causalities constraints through the relay and GBS, thereby the sums up-link transmission data rate for all the AGUs should not exceed the relay UAV attainable data rate and the relay UAV transmission data rate should not exceed the GBS attainable data-rate. Considering the presence of constraints (\ref{eq:12}) and (\ref{eq:13}), it can be observed that the objective function does not exhibit joint concavity for the variables $\mathcal{P}$, $\mathcal{X}$ and $\mathcal{Q}$. As a result, problem $\mathcal{P}1$ is classified as a non-convex problem, posing significant challenges for direct solutions. Therefore, we use the following approach to solve $\mathcal{P}1$.
\section{Proposed solution}
To tackle the challenges posed by the non-convexity of problem $\mathcal{P}1$, we adopt a strategy of decomposing it into two distinct sub-parts. Subsequently, we develop an iterative algorithm that utilizes the block coordinate descent technique. This approach enables us to obtain locally optimal solutions for each sub-part iteratively. Although block coordinates descent techniques have been used in other related works \cite{27b, 50}, the problems investigated in those works are distinct from our research problem.\par
By introducing a slack variable $\tilde{\mathcal{R}}  \triangleq  \{\tilde{R_u}, \forall u\}$, such that $\tilde{R_u}=R_u(1-\rho)$, the problem $\mathcal{P}1$ can be formulated as:
\begin{equation}
\mathcal{P}2: \begin{array}{c}
\mbox{max}\\
\tilde{\mathcal{R}}, \mathcal{X}, \mathcal{P}, \mathcal{Q}
\end{array}\frac{\theta}{U}\sum_{u}^{U} \log \left(\beta \frac{\tilde{R_u}}{\overline{r}_u}\right)
\nonumber
\end{equation}
\qquad  \qquad  \textbf{s. t.} \qquad (\ref{eq:11}),~(\ref{eq:13})~-~(\ref{eq:18})
\begin{eqnarray}
\label{eq:22}
R_u(1-\rho) & \geq & \tilde{R_u} \qquad \forall u \\
\label{eq:23}
R_r & \geq & \sum_u^U\tilde{R_u}    
\end{eqnarray}
By replacing the inequality in constraint (\ref{eq:11}) with equality, we get
\begin{equation}
\mathcal{P}3: \begin{array}{c}
\mbox{max}\\
\tilde{\mathcal{R}}, \mathcal{X}, \mathcal{P}, \mathcal{Q}
\end{array}\frac{\theta}{U}\sum_{u}^{U} \log \left(\beta \frac{\tilde{R_u}}{\overline{r}_u}\right)
\nonumber
\end{equation}
\qquad  \qquad \qquad \qquad \qquad \qquad \textbf{s. t.} \qquad (\ref{eq:13})~-~(\ref{eq:23})
\begin{equation}
\label{eq:24}
P_u^{out} = \rho, \quad \forall u
\end{equation}
\emph{\textbf{Theorem 1:}} Problem $\mathcal{P}2$ equals problem $\mathcal{P}3$. \\
\emph{\textbf{Proof:}} Without losing the optimality of $\mathcal{P}2$, the constraints on (\ref{eq:11}) can be changed by the constraints on (\ref{eq:24}). Moreover, if there is an AGU $u$ in the optimal solution of $\mathcal{P}2$ for which the constraint (\ref{eq:11}) is satisfied, whereas the subsequent constraint (\ref{eq:24}) is not fulfilled, then, we can increase $R_u$ and the remaining adjustable variables such that constraint (\ref{eq:24}) is satisfied without changing the value of the objective, as $P_u^{out}$ does not diminish with $R_u$. Consequently, within problem $\mathcal{P}2$, there exists an optimal solution that ensures the fulfillment of the constraints stated in  (\ref{eq:24}). Consequently, solving problem $\mathcal{P}2$ can be regarded as equivalent to solving problem $\mathcal{P}3$, thus concluding the proof.\par

To write $R_u$ in terms of $F^{-1}(\rho)$, put $P_u^{out} = \rho$ in (\ref{eq:6}) and rearrange it. Moreover, replacing constraint (\ref{eq:24}) by $R_u$ in $\mathcal{P}3$, we get the following equivalent problem.
\begin{equation}
\mathcal{P}4: \begin{array}{c}
\mbox{max}\\
\tilde{\mathcal{R}}, \mathcal{X}, \mathcal{P}, \mathcal{Q}
\end{array}\frac{\theta}{U}\sum_{u}^{U} \log \left(\beta \frac{\tilde{R_u}}{\overline{r}_u}\right)
\nonumber
\end{equation}
\qquad \qquad \quad \textbf{s. t.} (\ref{eq:14}) - (\ref{eq:23})
\begin{eqnarray}
\label{eq:25}
R_u & = & x_u \log_2\left(1 + \frac{F^{-1}(\rho) P_u \mu_0}{x_u (H_o)^2 + \parallel q_o - w_u \parallel^2}\right) \\
\label{eq:26}
R_b &\geq & \sum_u^U\tilde{R_u}
\end{eqnarray}
${F^{-1}(\cdot)}$ denotes the inverse function of $F(\cdot)$. 
\par
Note that the Marcum-Q function is difficult to compute, therefore, and as a result, there is no closed-form expression for calculating $F^{-1} (\rho)$. However, since $F(\cdot)$ is a non-decreasing function, therefore for $0 \leq \rho \leq 1$ $F^{-1} (\rho)$ is also a non-decreasing function. Given $\rho$, the value of $F^{-1} (\rho)$ can be calculated using a bisection numerical approach as used in \cite{51}.\par
Considering the block coordinate descent technique \cite{50}, we subsequently suggest a systematic and effective iterative algorithm for $\mathcal{P}4$. The primary idea is to split $\mathcal{P}4$ into two sub-categories; firstly power ($\mathcal{P}$) and bandwidth ($\mathcal{X}$) allocation with fixed UAVs locations and secondly optimizing the UAVs locations ($\mathcal{Q}$) for fixed power and bandwidth. Then, we optimize the two variable sets, \{$\mathcal{X}$, $\mathcal{P}$\} and $\mathcal{Q} = \{q_u, q_r\}$ interchangeably through the block coordinate descent method.
\subsection{Power ($\mathcal{P}$) and bandwidth ($\mathcal{X}$) allocation with fixed UAVs location}
\label{Agupowerband}
\begin{equation}
\mathcal{P}5: \begin{array}{c}
\mbox{max}\\
\tilde{\mathcal{R}}, \mathcal{X}, \mathcal{P}
\end{array}\frac{\theta}{U}\sum_{u}^{U} \log \left(\beta \frac{\tilde{R_u}}{\overline{r_u}}\right)
\nonumber
\end{equation}
\qquad \qquad \qquad \textbf{s. t.} \qquad (\ref{eq:14})~-~(\ref{eq:23}),(\ref{eq:25})~-~(\ref{eq:26})\;.\\
\par
Upon observation, it becomes evident that the objective function in terms of $\tilde{\mathcal{R}_u}$ exhibits concavity, while the constraint functions are all convex in $\mathcal{P}$ and $\mathcal{X}$. Therefore, $\mathcal{P}5$ is considered a standard convex optimization problem that can be effectively solved using various standard convex optimization techniques or through CVX \cite{53}.
\subsection{Observation and relay UAV position ($\mathcal{Q}$) optimization with fixed power and bandwidth}
For a fixed allocated bandwidth $\mathcal{X}$ and transmit power $\mathcal{P}$, the next challenge is to optimize the observation and relay UAVs positions i.e., $\mathcal{Q} =\{q_o, q_r\}$, elaborated as
\begin{equation}
\mathcal{P}6: \begin{array}{c}
\mbox{max}\\
\tilde{\mathcal{R}}, \mathcal{Q}
\end{array}\frac{\theta}{U}\sum_{u}^{U} \log \left(\beta \frac{\tilde{R_u}}{\overline{r_u}}\right)
\nonumber
\end{equation}
\qquad \qquad \qquad \textbf{s. t.} \qquad (\ref{eq:22}),~(\ref{eq:25}), and (\ref{eq:26})\\
\par
Problem $\mathcal{P}6$ can also be written by expanding the constraint equations as follows,
\begin{equation}
\mathcal{P}6: \begin{array}{c}
\mbox{max}\\
\tilde{\mathcal{R}}, \mathcal{Q}
\end{array}\frac{\theta}{U}\sum_{u}^{U} \log \left(\beta \frac{\tilde{R_u}}{\overline{r_u}}\right)
\nonumber
\end{equation}
\qquad \qquad \qquad \textbf{s. t.}
\begin{equation}
\label{eq:27}
x_u \log_2\left(1 + \frac{\mu_u}{x_u (H_o)^2+\parallel q_o - w_u \parallel^2}\right) (1-\rho)\geq \tilde{R_u}
\end{equation}
\begin{equation}
\label{eq:28}
\log_2\left(1 + \frac{\mu_{ob}}{(H_r - H_o)^2 + \parallel q_r - q_o \parallel^2}\right) \geq \sum_u^U\tilde{R_u}
\end{equation}
\begin{equation}
\label{eq:29}
\log_2\left(1 + \frac{\mu_{r}}{(H_b - H_r)^2 + \parallel w_b - q_r \parallel^2}\right) \geq \sum_u^U\tilde{R_u} \;,
\end{equation}
where $\mu_u = P_u \mu_0⁄x_u$, $\mu_{ob} = P_o \mu_0$, and $\mu_r = P_r \mu_0$ are constants with given $\mathcal{P}$ and $\mathcal{X}$. The problem $\mathcal{P}6$ is non-convex, however, successive convex approximation of constraints (\ref{eq:27}),~(\ref{eq:28}), and (\ref{eq:29})  through first-order Taylor approximation based iterative approach can make the problem tractable. That is the constraints (\ref{eq:27}),~(\ref{eq:28}), and (\ref{eq:29}) at a given local point $\mathcal{Q}^i$ (where $\mathcal{Q}^i= \{q_o^i, q_r^i\}$, at the $i$th iteration) are replaced by their convex approximations which make the overall problem convex and is solved. This whole process is continued iteratively until the algorithm converges. Therefore, we can utilize the first-order Taylor approximation to approximate the logarithm function of (\ref{eq:27}),~(\ref{eq:28}), and (\ref{eq:29}) and considering the lower bounds, we have 
\begin{eqnarray}
x_u \log_2\left(1 + \frac{\mu_u}{x_u (H_o)^2+\parallel q_o - w_u \parallel^2}\right)\geq R_u^{LB} \nonumber \\
\label{eq:30}
R_u^{LB} \triangleq x_u \left(c_u^i - D_u^i\right) \left(\parallel q_o - w_u \parallel^2 - \parallel q_o^i - w_u \parallel^2\right)\;,
\end{eqnarray}
where 
\begin{eqnarray}
D_u^i & = & \frac{\mu_u \log_2 e}{\left(H_o^2 + \parallel q_o^i - w_u \parallel^2\right)\left(H_o^2 + \parallel q_o^i - w_u \parallel^2 + \mu_u\right)} \nonumber \\
\label{eq:31}
C_u^i & = & \log_2 \left(1 + \frac{\mu_u}{H_o^2 + \parallel q_o^i - w_u \parallel^2} \right)\;.
\end{eqnarray}
For (\ref{eq:28})
\begin{equation}
\label{eq:32}
R_o = \log_2\left(1 + \frac{\mu_{ob}}{(H_r - H_o)^2 + \parallel q_r - q_o \parallel^2}\right)\geq R_o^{LB}
\end{equation}
\begin{equation}
\label{eq:33}
R_o^{LB} \triangleq \left( C_o^i - D_o^i\right) \left(\parallel q_r - q_o \parallel^2 - \parallel q_r^i - q_o^i \parallel^2\right)\;,
\end{equation}
where
\begin{eqnarray}
\nonumber
D_o^i = \frac{\mu_{ob} \log_2 e}{((H_r - H_o)^2 + \parallel q_r^i - q_o^i \parallel^2)((H_r - H_o)^2 + \parallel q_r^i - q_o^i \parallel^2 + \mu_{ob})}\;,
\end{eqnarray}
\begin{equation}
\label{eq:35}
C_o^i = \log_2\left(1 + \frac{\mu_{ob}}{(H_r - H_o)^2 + \parallel q_r^i - q_o^i \parallel^2}\right)\;.
\end{equation}
For (\ref{eq:29})
\begin{equation}
\label{eq:36}
R_r = \log_2\left(1 + \frac{\mu_r}{(H_b - H_r)^2 + \parallel w_b - q_r \parallel^2}\right)\geq R_r^{LB}
\end{equation}
\begin{equation}
\label{eq:37}
R_r^{LB} \triangleq \left( c_r^i - D_r^i\right) \left(\parallel q_r - q_b \parallel^2 - \parallel q_r^i - w_b^i\parallel^2\right)\;,
\end{equation}
where
\begin{eqnarray}
\nonumber
D_r^i = \frac{\mu_{r} \log_2 e}{((H_b - H_r)^2 + \parallel q_r^i - w_b^i \parallel^2)((H_b - H_r)^2 + \parallel q_r^i - w_b^i \parallel^2 + \mu_{r})}\;,
\end{eqnarray}
\begin{equation}
\label{eq:39}
C_r^i = \log_2\left(1 + \frac{\mu_{r}}{(H_b - H_r)^2 + \parallel q_o^i - w_b^i \parallel^2}\right)\;.
\end{equation}
For the given $\mathcal{Q}^{i} = \{q_o^i, q_r^i\}$, replacing the lower bounds by $R_u^{LB}$, $\forall u$, $R_o^{LB}$ and $R_r^{LB}$, in (\ref{eq:30}),~(\ref{eq:33}),~(\ref{eq:37}), $\mathcal{P}6$ is approximated as
\begin{equation}
\mathcal{P}7: \begin{array}{c}
\mbox{max}\\
\tilde{\mathcal{R}}, \mathcal{Q}
\end{array}\frac{\theta}{U}\sum_{u}^{U} \log \left(\beta \frac{\tilde{R_u}}{\overline{r_u}}\right)
\nonumber
\end{equation}
\qquad \qquad \qquad \textbf{s. t.}
\begin{eqnarray}
\label{eq:40}
  R_u^{LB} & \geq & \tilde{R_u}\;, \qquad \forall u \\
\label{eq:41}
R_o^{LB} & \geq & \sum_u^U \tilde{R_u}\;,\\
\label{eq:42}
R_r^{LB} & \geq & \sum_u^U \tilde{R_u}\;.
\end{eqnarray}
Constraints (\ref{eq:40}),~(\ref{eq:41}), and (\ref{eq:42}) are all concave with respect to $\mathcal{Q}$ and $\tilde{R_u}$ jointly, and therefore problem $\mathcal{P}7$ is a convex optimization problem. Standard convex optimization methods and solvers like CVX can be used to solve this problem within a polynomial-time complexity. It is important to note that the lower bounds specified in Equations (\ref{eq:30}), (\ref{eq:33}), and (\ref{eq:37}), can be easily validated. If the constraints (\ref{eq:40}), (\ref{eq:41}), and (\ref{eq:42}) in $\mathcal{P}7$ are satisfied, it can be inferred that the constraints defined in Equations (\ref{eq:27}), (\ref{eq:28}), and (\ref{eq:29}) in $\mathcal{P}6$ are also satisfied. However, the reverse is not necessarily true. Consequently, the feasible solution space of $\mathcal{P}7$ is a subspace of that of $\mathcal{P}6$, making the optimum solution of $\mathcal{P}7$ a lower bound to that of $\mathcal{P}6$. To solve the fundamental non-convex problem $\mathcal{P}6$, we adopt an iterative approach that involves optimizing $\mathcal{P}7$ iteratively using the specified local point $\mathcal{Q}^i$ at each iteration.\par
Drawing from the aforementioned outcomes, by utilizing the block coordinate descent method, we introduce an iterative algorithm to address problem $\mathcal{P}4$. This algorithm involves partitioning the optimization variables from the original problem $\mathcal{P}4$ into two blocks, namely $\mathcal{Q} = \{q_o, q_r\}$ and \{$\mathcal{X}$, $\mathcal{P}$\}. Subsequently, we iteratively optimize the transmit power and bandwidth allotment \{$\mathcal{X}$, $\mathcal{P}$\} as well as the UAV (observation and relay) locations $\mathcal{Q}$. In each iteration, we solve either problem $\mathcal{P}5$ or $\mathcal{P}7$ accordingly, and the resulting solution is then utilized as the input for the subsequent iteration. The specific steps of this algorithm are summarized in the PSCNs Algorithm 1 below.\par
\begin{algorithm}
\caption{PSCN Iterative Algorithm for $\mathcal{P}4$}
\label{algorithm1}
\begin{algorithmic}[1]
\State \textbf{Initialize:} $P_u^{\text{max}}$, $P_o^{\text{max}}$, $P_r^{\text{max}}$, $\mathcal{Q}^o$, $\rho$, $P_u^o$, $P_o^o$, and $P_r^o$
\State Let $i = 1$
\State Use bisection numerical technique to obtain $F^{-1}(\rho)$
\Repeat
\State Solve problem $\mathcal{P}5$ for given $\mathcal{Q}^o$ to get $x^{i+1}$, $P^{i+1}$
\State Resolve problem $\mathcal{P}7$ for the obtained $x^{i+1}$ and $P^{i+1}$ to get $\mathcal{Q}^{i+1}$
\State $i = i + 1$
\Until convergence
\end{algorithmic}
\end{algorithm}

In PSCNs Algorithm 1, it is important to note that during each iteration, problem $\mathcal{P}7$ is solved optimally using the given local point $\mathcal{Q}^i$. It is observed that the objective value of $\mathcal{P}7$ does not decrease as the iterations progress. Moreover, the objective value is always bounded above by a finite value. As a result, we can establish the guarantee of convergence for PSCN Algorithm 1. Furthermore,
both $\mathcal{P}5$ and $\mathcal{P}7$ are convex optimization problems, and Algorithm 1 exhibits polynomial complexity, as elucidated below. In line 5 of the algorithm, the resolution of $\mathcal{P}5$ entails solving a convex problem with $3U+2$ variables and $4U+5$ constraints, where U represents the number of users. The accuracy of this solution is guaranteed, and its computational complexity is polynomial, specifically $O\left(\sqrt{3U+2}(4U+5)\log \frac{1}{\mu}\right)$ where $\mu > 0$ is the accuracy tolerance of the solution \cite{53a}. In line 6, $\mathcal{P}7$ is addressed, which is another convex problem characterized by $U+4$ variables and $U+2$ constraints, with a computational complexity of $O\left(\sqrt{U+4}(U+2)\log \frac{1}{\mu}\right)$ \cite{53a}. When considering the overall complexity of Algorithm I, it can be expressed as $O\left(\sqrt{3U+2}(4U+5)\log \frac{1}{\mu} + \sqrt{U+4}(U+2)\log \frac{1}{\mu}\right)$. Disregarding constant terms, the complexity in terms of the number of users, number of constraints, and the accuracy tolerance simplifies to approximately $O(U^{3/2}\log \frac{1}{\mu})$.
\section{Simulation Results}
\label{simulations}
This section contains the performance evaluation of the presented solution via extensive simulations using Matlab R2022a. Moreover, we consider a Rician fading channel between the AGUs and the observation UAV channel, whereas, the channel between the observation UAV and the relay UAV, as well as between the relay UAV and the GBS, are dominated by the LoS links. Furthermore, a $500 \times 500~m^2$ sized geographic area is considered with the uniform deployment of the AGUs. The parameters used in video streaming utility function are $\beta = 100$, $\overline{r}_u$ = 1 Mbps and $\theta = 0.8$  for every AGU $n_{u}$ as mentioned in \cite{27c}. The simulation parameters are summarized in Table~\ref{table_2}, outlining the various parameters employed in the simulations. The GBS is positioned at $w_b = [-2500,~0]^T$, and the observation and relay UAVs maximum transmission power is taken as $0.1$ Watt.\par
\begin{table}[h!]
\caption{Parameters notation and definition}
\label{table_2}
\begin{center}
\begin{tabular}{|c|l|l|}
\hline
\textbf{Variable} & \textbf{Definition} & \textbf{Used} \\
\textbf{Name} & \textbf{Description} & \textbf{Value} \\
\hline
$H_o$	& Observation UAV height (m) & 100 m \\
\hline
$H_r$	& Relay UAV height (m)	& 100 m \\
\hline
$H_b$	& Ground  BS height	(m) & 20 m \\
\hline
$K_c$	& Rician fading factor	& 4 \\
\hline
$\rho$	& Target outage probability threshold & $10^{-2}$\\
\hline
 &	Maximum allowed transmission power for & \\
$P_o^{max}$ & observation UAV (Watts)	&   0.1Watt\\
\hline
& Maximum allowed transmission power for & \\
$P_r^{max}$ &  relay UAV (Watts) & 	0.1 Watt \\
\hline
& Maximum transmit power of the & \\
$P_u^{max}$ & ground users (Watts) & 0.2 Watt\\
\hline
$B$ & Signal bandwidth (Hz)	& 1MHz\\
\hline
$N_0$ & Noise power spectral density & $-170$ dBm / Hz \\
\hline
$\alpha_0$	& Channel power gain at 1m reference distance	& -60~dB\\
\hline
\end{tabular}
\end{center}
\end{table}
\begin{figure}[h!]
\centering
\includegraphics[width=0.9\linewidth]{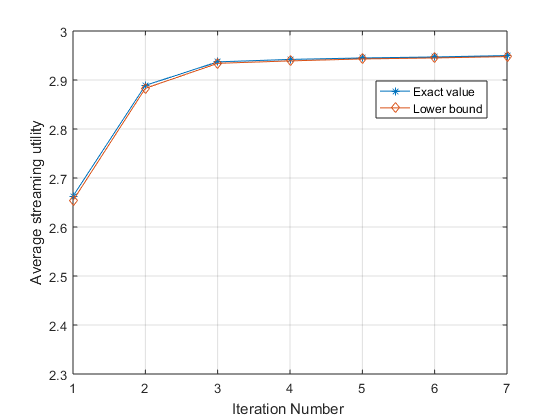}
\caption{Proposed iterative algorithm-1 convergence curve}
\label{fig2}
\end{figure}
Primarily, we consider the convergence of the presented technique. In the suggested methodology, the observation UAV and relay UAV initial locations are considered as $q_o$ and $q_r$, respectively. Fig.~\ref{fig2} shows the convergence of PSCNs Algorithm with $U = 30$ and $P_u^{max} = P_o^{max} =  P_r^{max} = 0.1$ Watt. More specifically, we compare the lower bound value of the objective function obtained through the PSCNs Algorithm 1, with the actual value of the average streaming video utility computed using ($\mathcal{P}1$). Fig.~\ref{fig2} clearly shows that the two curves are overlapping, which clarifies the fact that the lower bound of the video streaming utility obtained using PSCNs Algorithm 1 is tight. In addition, Fig. \ref{fig2} also depicts that the presented technique converges fast, which validates the efficiency of the proposed technique.\par
\begin{figure}[h!]
\centering
\includegraphics[width=0.9\linewidth]{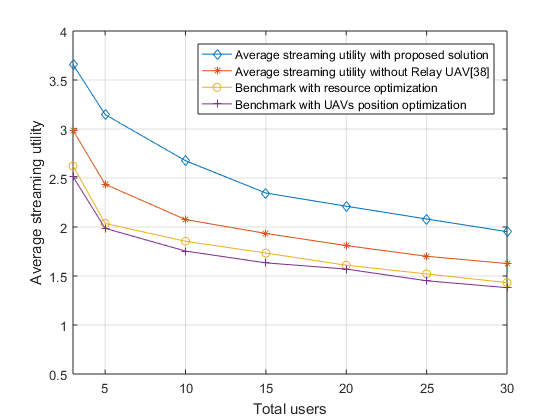}
\caption{Average streaming utility vs the number of AGUs}
\label{fig3}
\end{figure}
The performance metric employed in this study is the average video streaming utility. To assess the efficacy of our proposed solution, we conduct a comparison with four benchmark schemes: (1) a benchmark scheme involving resource optimization, (2) a benchmark scheme focusing on UAV position optimization, (3) the average streaming utility with relay UAVs, and (4) the average streaming utility without relay UAVs. In the benchmark with resource optimization, we maintain the UAV position as the initial position in Algorithm 1, and subsequently optimize the allocation of transmission bandwidth and transmit power through iterative solutions to problem $\mathcal{P}5$.

In the benchmark with UAVs position optimization for fixed power and bandwidth allocation, we iteratively solve problem $\mathcal{P}7$ to obtain an optimal solution. These benchmark schemes serve as reference points for comparison, allowing us to assess the performance of the presented solution. In the calculation of average streaming utility, we consider the approach as proposed in \cite{27b} and \cite{55} respectively, with and without a relay UAV.
 Fig.~\ref{fig3} illustrates the effect of the AGUs count on the performance for the above-mentioned techniques with the maximum power of observation UAV, relay UAV, and AGUs, i.e., $ P_o^{max} =  P_r^{max} = 0.1 W$ and $P_u^{max} = 0.2W$. The result reveals a decrease in the average streaming utility as the number of AGUs increases, which is expected as more AGUs will compete for the limited communication resources, especially when U is large. However, our presented solution outperforms the other schemes as the performance is improving with an increase in the number of AGUs, as shown in Fig.~\ref{fig3}.\par
\begin{figure}[h!]
\centering
\includegraphics[width=0.9\linewidth]{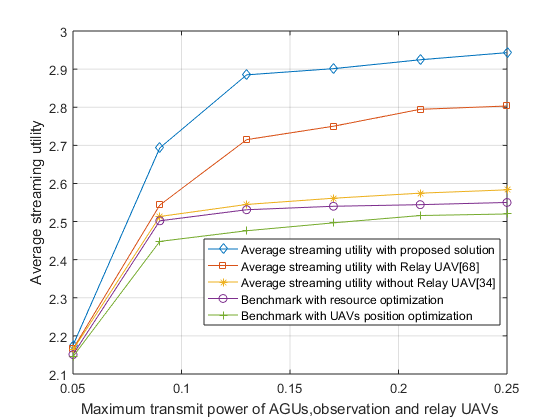}
\caption{Average Steaming Utility vs $P_u^{max}$, $P_o^{max}$ and $P_r^{max}$}
\label{fig4}
\end{figure}
In Fig.~\ref{fig4}, the average video streaming utility is compared while varying the maximum transmit powers $P_u^{max}$, $P_o^{max}$ and $P_r^{max}$ of the AGUs, observation, and relay UAVs, respectively, with a total of $U = 10$ users. As anticipated, an increase in $P_u^{max}$, $P_o^{max}$, and $P_r^{max}$ results in an improved average video streaming utility for all four techniques considered. This is due to the higher transmission power, which allows increased transmission rates for each AGU, resulting in improved video streaming performance. Moreover, as obvious from Fig.~\ref{fig4}, our proposed solution still has improved performance in terms of average video streaming utility among all. The saturation of the average streaming utility with higher $P_u^{max}$, $P_o^{max}$ and $P_r^{max}$ is also observed. This is owing to the streaming utility function definition taking into account the value of transmission rate that provides a validation for the perception that user utility saturation occurs with a rise in the rate of transmission.\par
\begin{figure}[h!]
\centering
\includegraphics[width=0.9\linewidth]{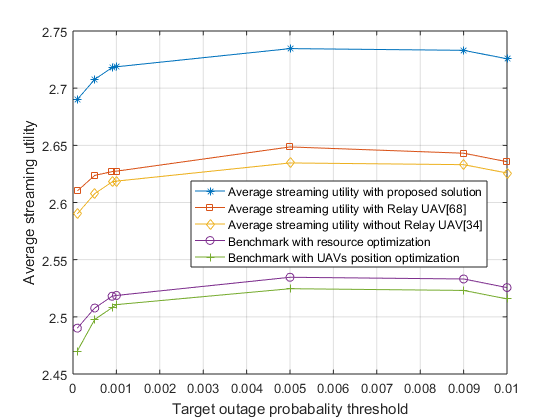}
\caption{Average streaming utility vs $\rho$}
\label{fig5}
\end{figure}
Fig.~\ref{fig5} illustrates that the average streaming utility experiences a rise followed by a fall as the target outage probability threshold $\rho$ increases. This trend is particularly noticeable when the value of $\rho$ is small. An increase in $\rho$ leads to an increase in the transmission rate, resulting in an improvement in the streaming utility. However, as the outage probability threshold becomes large due to the effects of fading channels, the quality of experience (QoE) suffers, leading to video play-out stalling. This trade-off between streaming utility and QoE is clearly illustrated in Fig.~\ref{fig5}.\par
\begin{figure}[h!]
\centering
\includegraphics[width=0.9\linewidth]{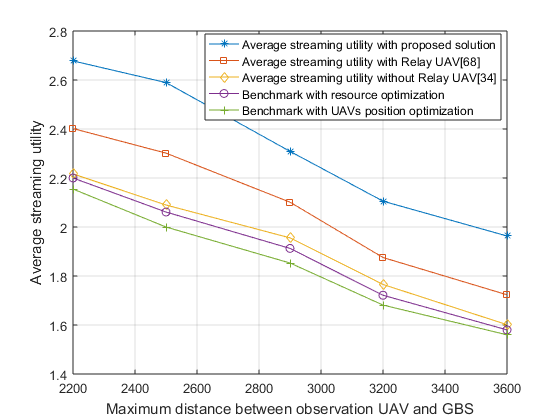}
\caption{Average streaming utility vs D}
\label{fig6}
\end{figure}
Fig.~\ref{fig6} illustrates the benefits of optimal relay placement by varying the network size (D) from 1500m to 3200m while keeping the observation area radius fixed at 500m and the maximum power for both observation and relay at 0.1W. As the network size increases, the observation area moves further away from the GBS. However, it is clear from Fig.~\ref{fig6} that our solution outperforms the remaining solutions.\par
\begin{figure*}[h!]
\centering
\includegraphics[width=0.9\linewidth]{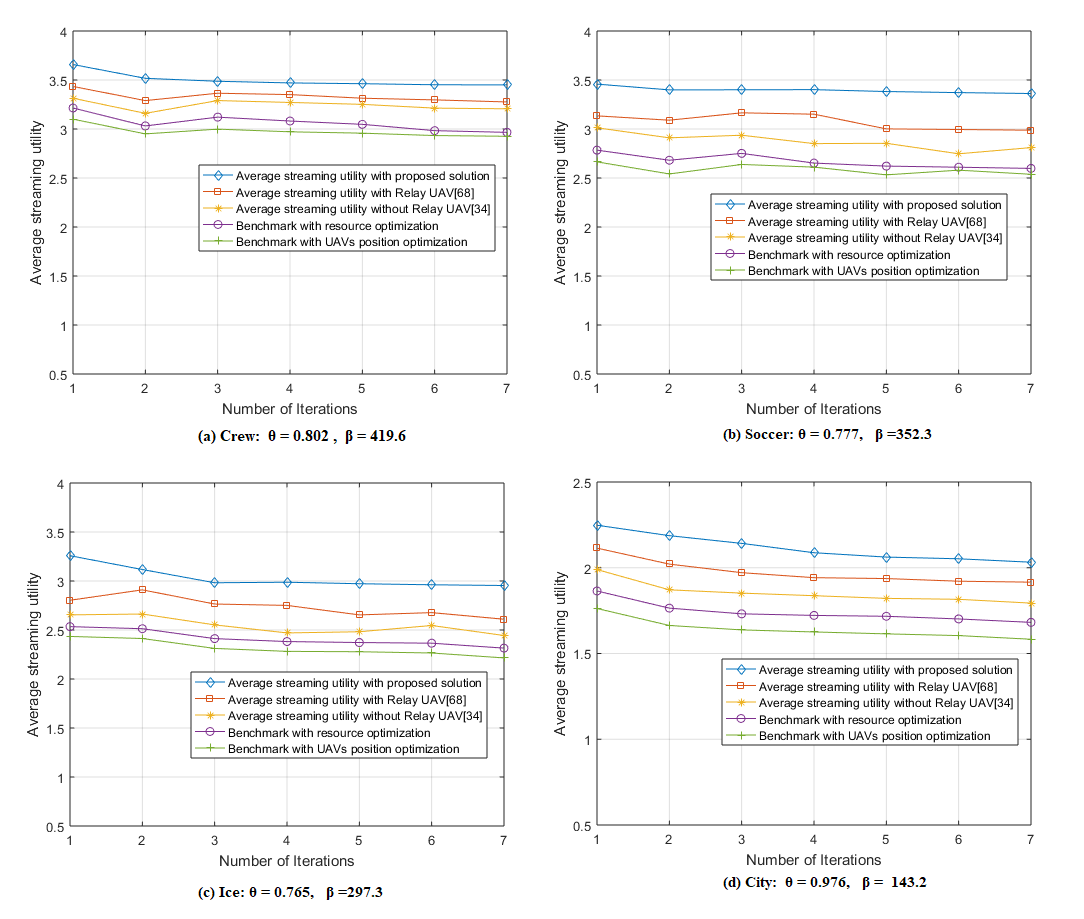}
\caption{Proposed method with different video scenarios and controlling constant parameters $\theta$, $\beta$}
\label{fig7}
\end{figure*}
Fig.~\ref{fig7} evaluates the utility of video streaming in various video scenarios to better understand the performance of our proposed method in real-world video scenarios. Each of these videos has different QoE parameters \cite{54}. Furthermore, we evaluate the effectiveness of our proposed approach across these video scenarios by employing various comparative criteria, providing clear evidence of its superiority over competing methods.\par 

\section{Conclusion}
\label{conclusion}
In this work, we focused on the joint optimization of observation and relay UAV positions, as well as bandwidth and transmit power allocation in UAV-assisted PSCNs, specifically for uplink video stream transmission. We proposed a unified design approach to maximize the average streaming quality for all users, taking into account the impact of fading channels in PSCNs. By leveraging the block coordinate descent and successive convex approximation techniques, we developed an efficient iterative algorithm and analyzed its convergence properties. Our simulation results demonstrated that the proposed approach can significantly enhance the maximum average video streaming quality in PSCNs.



\begin{IEEEbiography}[{\includegraphics[width=1in,height=1.25in,clip,keepaspectratio]{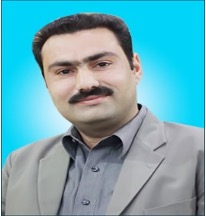}}]{\textbf{Naveed Khan}} received his Bachelor’s degree in Telecommunication engineering from University of engineering and technology (UET), Peshawar and Master’s degree in Electronics Engineering from International Islamic University Islamabad (IIUI) in 2009 and 2013, respectively. He is a PhD student at COMSATS University, Islamabad, Wah Campus, Pakistan. He is currently an Assistant Professor in the Department of Electrical Engineering at Abasyn University Peshawar, Pakistan. His research interests include unmanned aerial vehicle communications, image processing, and multimedia communications. 
\end{IEEEbiography}

\begin{IEEEbiography}[{\includegraphics[width=1in,height=1.25in,clip,keepaspectratio]{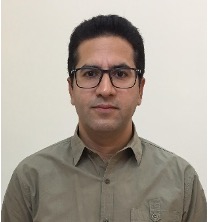}}]{\textbf{Ayaz Ahmad}} (S’08, M’15, SM’16) is currently an Associate Professor with the Department of Electrical and Computer Engineering, COMSATS University Islamabad – Wah Campus, Pakistan. Prior to that, he was an Assistant Professor at the same university. He received a B.Sc. degree in electrical engineering from the University of Engineering and Technology, Peshawar, Pakistan, in 2006, and the M.S. and Ph.D. degrees in Wireless Communication from Ecole Superieure d'Electricite (Supelec), Gif-sur-Yvette, France, in 2008 and 2011, respectively. From 2006 to 2007, he was a Faculty Member with the Department of Electrical Engineering, FAST-NUCES, Peshawar, Pakistan. He is the recipient of the best research paper award from the Higher Education Commission, Pakistan for the years 2015-2016, the National Research Productivity Award from Pakistan Council of Science and Technology (PSCT) in 2017, the best research paper award in IEEE IEMCON 2018 held in Canada, and Publon Top Peer Reviewer award in 2019. He has several years of research experience and has authored or co-authored several scientific publications in various refereed international journals and conferences. He has also authored or co-authored several book chapters and is the leading Co-Editor of the book Smart Grid as a Solution for Renewable and Efficient Energy published in 2016. He is Associate Editor with IEEE Access, Human-centric Computing and Information Sciences, PeerJ Computer Science, and Frontiers in Smart Grids. He regularly serves as a TPC member for several international conferences, including IEEE GLOBECOM, IEEE ICC, and IEEE PIMRC, and as a reviewer for several renowned international journals. He is a Senior Member IEEE, and Associate Fellow of Advance HE, UK. He is a member of the IEEE Communication Society. His research interests include resource allocation in wireless communication systems, energy management in smart grids, and the application of optimization methods to engineering problems. 
\end{IEEEbiography}

\begin{IEEEbiography}[{\includegraphics[width=1in,height=1.25in,clip,keepaspectratio]{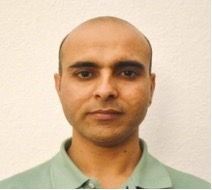}}]{\textbf{Abdul Wakeel}} received his Ph.D. from Jacobs University Bremen, Bremen, Germany, in 2016. He joined the National University of Sciences and Technology (NUST) in 2016 as an Assistant Professor. His research areas are PAPR reduction for OFDM-based SISO/ MIMO systems, Error Control Coding, Signal Processing, Active noise control,  Physical layer security, and wireless communications.
\end{IEEEbiography}

\begin{IEEEbiography}[{\includegraphics[width=1in,height=1.25in,clip,keepaspectratio]{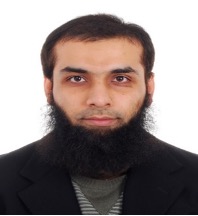}}]{\textbf{Zeeshan Kaleem}} (Senior Member, IEEE) is serving as an Associate Professor in the Electrical and Computer Engineering Department, COMSATS University Islamabad, Wah Campus. He received his BS from UET Peshawar in 2007. He received MS and Ph.D. in Electronics Engineering from South Korea in 2010 and 2016, respectively. Dr. Zeeshan consecutively received the National Research Productivity Award (RPA) awards from the Pakistan Council of Science and Technology (PSCT) in 2017 and 2018. We won the Runner-up Award in the National Hackathon 23 competition for Project to develop a drone detection system.  He also won the Higher Education Commission (HEC) Best Innovator Award in 2017. He is the recipient of the 2021 Top Reviewer Recognition Award for IEEE TRANSACTIONS on VEHICULAR TECHNOLOGY. He has published around 70 technical Journal papers, Book Chapters, and conference papers in reputable journals/venues and holds 21 US and Korean Patents, and got several research grants. He is a co-recipient of the best research proposal award from SK Telecom, Korea. He is currently serving as Technical Editor of several prestigious Journals/Magazines like IEEE Transactions on Vehicular Technology, Elsevier Computer and Electrical Engineering, Springer Human-centric Computing and Information Sciences, and Journal of Information Processing Systems, Frontiers in Communications and Networks. He has served/serving as Guest Editor for special issues in IEEE Wireless Communications, IEEE Communications Magazine, IEEE Access, Sensors, IEEE/KICS Journal of Communications and Networks, and Physical Communications, and also served as TPC Member for IEEE Globecom, IEEE VTC, IEEE ICC, and IEEE PIMRC.
\end{IEEEbiography}

\begin{IEEEbiography}[{\includegraphics[width=1in,height=1.25in,clip,keepaspectratio]{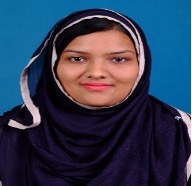}}]{\textbf{Bushra Rashid}} is currently serving as an Assistant Professor in the biomedical engineering department, King Faisal University, KSA. She did B.S. in Electrical Engineering from Wah Engineering College, Wah Cantt, Pakistan in 2013 and M.S. and Ph.D. in Electrical Engineering from Comsats University, Islamabad, Wah Campus, Pakistan on topics related to multimedia, signal processing, data security, and AI. 
\end{IEEEbiography}

\begin{IEEEbiography}[{\includegraphics[width=1in,height=1.25in,clip,keepaspectratio]{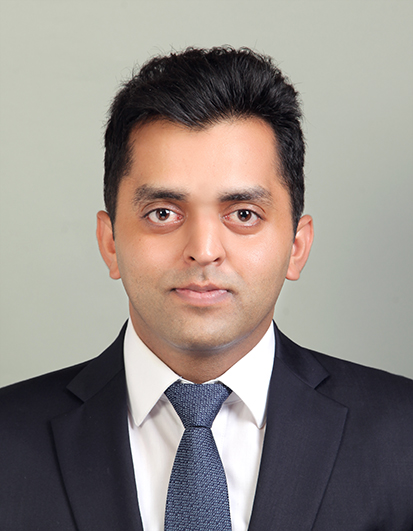}}]{\textbf{Waqas Khalid}} received a B.S. degree in Electronics Engineering from GIK Institute of Engineering Sciences and Technology, KPK, Pakistan, in 2011. He received M.S. and Ph.D. degrees in Information and Communication Engineering from Inha University, Incheon, South Korea, and Yeungnam University, Gyeongsan, South Korea, in 2016 and 2019, respectively. He is currently working as a research professor at the Institute of Industrial Technology, Korea University, Sejong, South Korea. His areas of interest include physical layer modeling, signal processing for wireless communications, and emerging solutions and technologies for 5G/6G networks such as reconfigurable intelligent surfaces, energy harvesting, physical layer security, NOMA, cognitive radio, UAVs, and IoTs.
\end{IEEEbiography}

\EOD
\end{document}